\documentclass[12pt]{article}
\usepackage{amsmath,amssymb}

\begin{document}

\begin{center}

{\Large \bf
Relativistic action at a distance and fields}
\vskip .6in

Domingo J. Louis-Martinez
\vskip .2in

Department of Physics and Astronomy,\\ University of British
Columbia\\Vancouver, Canada, V6T 1Z1 

martinez@physics.ubc.ca

\end{center}
 
\vskip 3cm
 
\begin{abstract}
After a brief review of the field formulations and the relativistic non-instantaneous action-at-a-distance
formulations of some well known classical theories, we study Rivacoba's generalization of a theory with a linearly rising
potential as a relativistic non-instantaneous action-at-a-distance theory. For this case we construct the corresponding field
theory, which turns out to coincide with a model proposed by Kiskis to describe strong interactions. We construct the action
functional for this field theory.
Although this model belongs to the class of Lagrangian theories with higher derivatives, it is shown that its action-at-a-distance 
counterpart has a much simpler
form. A proposal to construct an action-at-a-distance description of Yang-Mills interactions is also presented.
\end{abstract}

PCAS No. 03.30.+p, 03.50.-z.

The concept of action at a distance was introduced by Isaac Newton to describe gravitational interactions. It became the dominant paradigm
in physics until approximately the last quarter of the XIXth century. Relativistic action at a distance became a topic of research after the works of 
Wheeler and Feynman \cite{wheeler} on classical electrodynamics. Several non-instantaneous action-at-a-distance theories, modeled after
Wheeler-Feynman electrodynamics have been studied\cite{dettman}-\cite{friedman}. Instantaneous action-at-a-distance formulations have been investigated using a
variety of approaches\cite{bel}-\cite{longhi}.

Within the action-at-a-distance framework, great effort has been devoted to the investigation of relativistic bound states.
New results for the relativistic two-body problem have been obtained recently by Droz-Vincent \cite{droz-vincent1} using the a priori Hamiltonian
formalism of predictive relativistic dynamics. 
The relativistic two-body problem has also been considered recently in \cite{uri, luca}.

In this paper, after a short review of the relation between the action-at-a-distance and field formulations of well known 
classical theories, we focus on the study of models of vector confinement, both in their action-at-a-distance and field formulations.
We find a correspondence between Rivacoba's\cite{rivacoba} action at a distance theory and a field theory proposed by Kiskis\cite{kiskis} in the
70's as a model of vector confinement. We construct the action functional for this field theory. It is shown that the action-at-a-distance
formulation in this case has a simpler form than its field theory counterpart, which is a Lagrangian field theory with higher derivatives. 
We also present a proposal to construct an action-at-a-distance formulation in the case of Yang-Mills interactions. As in the case of 
electrodynamics, it is expected that the two-body problem becomes more tractable in this formulation.  

Classical electrodynamics is usually presented as a field theory. For example, it is well known that this theory can
be described by the action functional\cite{barut}:

\begin{equation}
S = -mc\int ds (\dot{z}^{2})^{\frac{1}{2}} - \frac{e}{c}\int ds A_{\mu} \dot{z}^{\mu} - \frac{1}{16\pi c}\int d^{4}x F_{\mu\nu}F^{\mu\nu},
\label{1}
\end{equation}

\noindent where $s=c\tau$, $\tau$ is the proper time of the charged particle, $c$ is the speed of light in vacuum and
$m$ and $e$ are the mass and electric charge of the particle. In (\ref{1}) $A^{\mu}$ ($\mu=0,1,2,3$) are the components 
of the 4-vector potential of the electromagnetic field and

\begin{equation}
F^{\mu\nu} = \partial^{\mu}A^{\nu} - \partial^{\nu}A^{\mu}.
\label{2}
\end{equation}

The world line of the particle in Minkowski spacetime is described by the 4-vector $z^{\mu}(s)$. We denote the 4-vector velocity
$\dot{z}^{\mu} = \frac{d z^{\mu}}{d s}$.

The equations of the theory take the form:

\begin{equation}
m \ddot{z}^{\mu} = \frac{e}{c^{2}} F^{\mu\nu} \dot{z}_{\nu},
\label{3}
\end{equation}

\begin{equation}
\square A^{\mu} - \partial^{\mu} \left(\partial_{\nu} A^{\nu}\right) = \frac{4\pi}{c} j^{\mu},
\label{4}
\end{equation}

\noindent where $j^{\mu}$ is the current 4-vector associated to the charged particle:

\begin{equation}
j^{\mu} = ce\int ds \delta^{4} \left(x - z(s)\right) \dot{z}^{\mu}(s).
\label{5}
\end{equation}

It is known that classical electrodynamics admits an action at a distance formulation\cite{wheeler}, \cite{barut}-\cite{hoyle}. In this
formulation the interactions are not mediated by fields. 

The action-at-a-distance theory of electrodynamics is described by the action functional:

\begin{equation}
S = - \sum\limits^{}_{i} m_{i} c \int d s_{i} \left(\dot{z}_{i}^{2}\right)^{\frac{1}{2}} 
- \frac{1}{2 c}\sum\limits^{}_{i}  \sum\limits^{}_{j\neq i}
e_{i} e_{j}
\int \int d s_{i} d s_{j}
\delta\left(\left(z_{i} - z_{j}\right)^{2}\right) \left(\dot{z}_{i} \dot{z}_{j}\right).
\label{8}
\end{equation}

In this formulation the charged particles obey the equations of motion:

\begin{equation}
m_{i} \ddot{z}_{i}^{\mu} = \frac{e_{i}}{c^{2}} F_{i}^{\mu\nu} \dot{z}_{i\nu},
\label{6}
\end{equation}

\noindent where $i = 1,2,..., N$ ($N$ is the total number of charged particles) and $F_{i}^{\mu\nu}$ are antisymmetric tensors given by
the expressions:

\begin{eqnarray}
F^{\mu\nu}_{i} & = & \sum\limits^{}_{j\neq i} e_{j} \int d s_{j}
\frac{\delta\left(\left(z_{i} - z_{j}\right)^{2}\right)}
{\left(\dot{z}_{j} (z_{i} - z_{j})\right)^{2}} 
\left(\left((z^{\mu}_{i} - z^{\mu}_{j})\dot{z}^{\nu}_{j} -
\dot{z}^{\mu}_{j} (z^{\nu}_{i} - z^{\nu}_{j})\right)
\left(1 - (\ddot{z}_{j} (z_{i} - z_{j}))\right) \right. \nonumber\\
& & + \left. \left((z^{\mu}_{i} - z^{\mu}_{j})\ddot{z}^{\nu}_{j} -
\ddot{z}^{\mu}_{j} (z^{\nu}_{i} - z^{\nu}_{j})\right)
\left(\dot{z}_{j} (z_{i} - z_{j})\right)\right)
\label{7}
\end{eqnarray}

The Dirac delta function in (\ref{7}) accounts for the interactions propagating at the speed of light forward and backward
in time.

The description given above is an example of a non-instantaneous action-at-a-distance theory.

The equations of motion (\ref{6}, \ref{7}) admit exact circular orbits solutions for any number of particles\cite{schild, domingo}.

Similarly, it is known that if the interactions between particles are mediated by a scalar the corresponding field theory
can be described by the action functional\cite{anderson}:

\begin{equation}
S = -c\int ds \left(m + \frac{g \phi}{c^{2}}\right)(\dot{z}^{2})^{\frac{1}{2}} - \frac{1}{8\pi c}\int d^{4}x \partial_{\alpha} \phi \partial^{\alpha} \phi,
\label{11}
\end{equation}

\noindent or equivalently by the equations:

\begin{equation}
\frac{d}{ds} \left(\left(m  +  \frac{g \phi}{c^{2}}\right) \dot{z}^{\mu}\right) = \frac{g}{c^{2}} \partial^{\mu} \phi,
\label{12}
\end{equation}

\begin{equation}
\square \phi = 4\pi g\int ds \delta^{4} \left(x - z(s)\right) 
\label{13}
\end{equation}

This theory too admits an action-at-a-distance formulation described by the action functional\cite{anderson}:

\begin{equation}
S = - \sum\limits^{}_{i} m_{i} c \int d s_{i} \left(\dot{z}_{i}^{2}\right)^{\frac{1}{2}} 
- \frac{1}{2 c}\sum\limits^{}_{i}  \sum\limits^{}_{j\neq i}
g_{i} g_{j}
\int \int d s_{i} d s_{j}
\delta\left(\left(z_{i} - z_{j}\right)^{2}\right) (\dot{z}_{i}^{2})^{\frac{1}{2}}(\dot{z}_{j}^{2})^{\frac{1}{2}}.
\label{14}
\end{equation}

This is a relativistic action-at-a-distance theory of particles with variable rest masses\cite{vonBaeyer, domingo2, domingo3}:

\begin{equation}
\bar{m}_{i} = m_{i} + \frac{g_{i}}{c^2} \sum\limits^{}_{j\neq i} g_{j}
\int d s_{j} \delta\left(\left(z_{i} - z_{j}\right)^{2}\right).
\label{15}
\end{equation}

The particles obey the equations of motion\cite{domingo2, domingo3}:

\begin{equation}
\bar{m}_{i} \ddot{z}_{i}^{\mu} = \frac{g_{i}}{c^{2}} \Gamma_{i\alpha\beta}^{\mu} \dot{z}_{i}^{\alpha}\dot{z}_{i}^{\beta},
\label{16}
\end{equation}

\noindent where:

\begin{eqnarray}
\Gamma^{\mu}_{i\alpha\beta} & = &
\sum\limits^{}_{j\neq i} g_{j} \int d s_{j}
\frac{\delta\left((z_{i} - z_{j})^{2}\right)}
{\left(\dot{z}_{j}(z_{i} - z_{j})\right)^{2}} \nonumber\\
& & \left[\left((z^{\mu}_{i} - z^{\mu}_{j})\eta_{\alpha\beta} -
\frac{1}{2}\left(\delta^{\mu}_{\alpha}(z_{i\beta} - z_{j\beta})
+ \delta^{\mu}_{\beta}(z_{i\alpha} - z_{j\alpha})\right)\right)
\left(1 - (\ddot{z}_{j} (z_{i} - z_{j}))\right)\right. \nonumber\\
& & -\left. \left(\dot{z}^{\mu}_{j}\eta_{\alpha\beta} -
\frac{1}{2}(\delta^{\mu}_{\alpha}\dot{z}_{j\beta}
+ \delta^{\mu}_{\beta}\dot{z}_{j\alpha})\right) \left(\dot{z}_{j}(z_{i} - z_{j})\right)
\right]
\label{18}
\end{eqnarray}

In the context of the action-at-a-distance formulations, in \cite{domingo2} the following action functional,
describing a system of $N$ interacting relativistic particles, was investigated:

\begin{equation}
S = - \sum\limits^{}_{i} m_{i}c\int d\lambda_{i} \zeta_{i}^{\frac{1}{2}}  
- \frac{1}{2 c} \sum\limits^{}_{i}  \sum\limits^{}_{j\neq i} g_{i} g_{j}
\int \int d\lambda_{i} d\lambda_{j}  \delta\left(\rho_{ij}\right)
F\left(\xi_{ij}, \gamma_{ij},\gamma_{ji}, \zeta_{i}, \zeta_{j} \right)
\label{19}
\end{equation}

In (\ref{19}) the $\lambda_{i}$ ($i=1,...,N$) are Poincare invariant parameters labeling the events along the world lines
of the particles and $\dot{z}_{i}^{\mu}=\frac{d z^{\mu}}{d \lambda_{i}}$. The metric tensor $\eta_{\mu\nu}= diag(+1,-1,-1,-1)$
and the Poincare invariants $\zeta_{i}$, $\xi_{ij}$, $\gamma_{ij}$ and $\rho_{ij}$ are defined as follows:

\begin{equation}
\zeta_{i} = \dot{z}_{i}^{2}
\label{20}
\end{equation}

\begin{equation}
\xi_{i j}= \left(\dot{z}_{i} \dot{z}_{j}\right) 
\label{21}
\end{equation}

\begin{equation}
\gamma_{i j}= \left(\dot{z}_{i} (z_{j} - z_{i})\right) 
\label{22}
\end{equation}

\begin{equation}
\rho_{i j}= \left( z_{i} - z_{j}\right)^{2} 
\label{23}
\end{equation}

The action functional (\ref{19}) is the most general action depending only on the 4-vector velocities and the relative positions
of the particles. It does not depend on the 4-vector accelerations or on higher derivatives. In (\ref{19}) it is assumed that the 
particles interact pairwise at a distance in flat spacetime and that the interactions propagate at the speed of light in vacuum.

In \cite{domingo2} it was shown that if ({\it i}) the Poincare invariant parameters $\lambda_{i}$ are identified with the proper times
of the particles ($\lambda_{i} = s_{i} = c \tau_{i}$) and ({\it ii}) all the masses are assumed to be scalars under Poincare transformations,
then the most general expression for $F\left(\xi_{ij}, \gamma_{ij},\gamma_{ji}, \zeta_{i}, \zeta_{j} \right)$ that satisfies these conditions is:

\begin{equation}
F = \alpha \gamma_{ij} \gamma_{ji} + \beta \xi_{ij}
+ \gamma \zeta_{i}^{\frac{1}{2}} \zeta_{j}^{\frac{1}{2}},
\label{24}
\end{equation}

\noindent where $\alpha$, $\beta$ and $\gamma$ are constants.

Let us consider the following three cases: (a) $\alpha \ne 0, \beta=\gamma=0$, (b) $\alpha=0, \beta \ne 0 ,\gamma=0$, 
(c) $\alpha =\beta= 0, \gamma \ne 0$. It is not difficult to see that case (b) corresponds to the functional (\ref{8}),
which describes electrodynamics. Case (c) corresponds to the action functional (\ref{14}), which describes scalar interactions
between the particles. As noted above, these two cases admit formulations as field theories.

The case (a) ($\alpha \ne 0, \beta=\gamma=0$) coincides with a theory proposed in \cite{rivacoba} by Rivacoba. This 
action-at-a-distance model is described by the action functional:

\begin{equation}
S = - \sum\limits^{}_{i} m_{i} c \int d s_{i} \left(\dot{z}_{i}^{2}\right)^{\frac{1}{2}} 
- \frac{\alpha}{2 c}\sum\limits^{}_{i}  \sum\limits^{}_{j\neq i}
g_{i} g_{j}
\int \int d s_{i} d s_{j}
\delta\left(\left(z_{i} - z_{j}\right)^{2}\right) \left(\dot{z}_{i} (z_{j} - z_{i})\right) \left(\dot{z}_{j} (z_{i} - z_{j})\right).
\label{25}
\end{equation}

The relativistic equations of motion for the system consisting of $N$ particles take the form:

\begin{equation}
m_{i} \ddot{z}_{i}^{\mu} = \frac{g_{i}}{c^{2}} F_{i}^{\mu\nu} \dot{z}_{i\nu},
\label{26}
\end{equation}

\noindent where,

\begin{equation}
F^{\mu\nu}_{i} = \alpha \sum\limits^{}_{j\neq i} g_{j} \int d s_{j}
\delta\left(\left(z_{i} - z_{j}\right)^{2}\right)
\left((z^{\mu}_{i} - z^{\mu}_{j})\dot{z}^{\nu}_{j} -
\dot{z}^{\mu}_{j} (z^{\nu}_{i} - z^{\nu}_{j})\right)
\label{27}
\end{equation}

In the non-relativistic approximation this theory is derived from the Lagrangian \cite{rivacoba, domingo3}:

\begin{equation}
L  =  \frac{1}{2} \sum\limits^{}_{i} m_{i} v_{i}^{2}
+ \frac{\alpha}{2} \sum\limits^{}_{i} \sum\limits^{}_{j \neq i}
g_{i} g_{j} r_{ij}
\label{28}
\end{equation}

\noindent where, $\vec{r}_{ij} = \vec{r}_{i} - \vec{r}_{j}$ is the relative position of
particle $i$ with respect to particle $j$ and $\vec{v}_{i}$ is the velocity of the i$^{th}$ particle.
The potential energy depends linearly on the distances $r_{ij} = |\vec{r}_{ij}|$ between particles (a linear potential).

In the theory of strong interactions confinement is associated with a linearly rising potential\cite{allen,sazdjian}. 
The relativistic non-instantaneous action-at-a-distance formulation
of \cite{domingo2} predicts that a linearly rising potential is purely a vector potential\cite{domingo3} (vector confinement). This is in agreement
with the analysis given in \cite{allenolsson} concluding that scalar confinement appears to be ruled out by experiments.

In the semirelativistic approximation (up to terms of second order ($\frac{v^{2}}{c^{2}}$)) the Lagrangian of the
theory takes the form\cite{domingo3}:

\begin{eqnarray}
L & = & \sum\limits^{}_{i}
\frac{m_{i} v_{i}^{2}}{2}
+ \sum\limits^{}_{i}
\frac{m_{i} v_{i}^{4}}{8 c^{2}}
+ \frac{\alpha}{2} \sum\limits^{}_{i} \sum\limits^{}_{j \neq i} g_{i} g_{j} r_{ij} \nonumber\\
& & + \frac{\alpha}{4 c^{2}}
\sum\limits^{}_{i} \sum\limits^{}_{j \neq i} g_{i} g_{j} r_{ij}
\left(- (\vec{v}_{i} \vec{v}_{j}) 
+ (\vec{n}_{ij} \vec{v}_{i})(\vec{n}_{ij} \vec{v}_{j})\right)
\label{29}
\end{eqnarray}

\noindent where $\vec{n}_{ij} \equiv \frac{\vec{r}_{ij}}{r_{ij}}$.

We have shown that the theory described by the action (\ref{25}) (and the equations of motion (\ref{26},\ref{27})) can be
obtained from general considerations and assumptions within the action-at-a-distance paradigm. Does this theory have a formulation as a
field theory? To answer this question, let us first notice that the antisymmetric tensors $F_{i}^{\mu\nu}$ given by (\ref{27}) obey the
Bianchi identities:

\begin{equation}
\frac{\partial F_{i}^{\alpha\beta}}{\partial z_{i\gamma}} + \frac{\partial F_{i}^{\beta\gamma}}{\partial z_{i\alpha}} 
+ \frac{\partial F_{i}^{\gamma\alpha}}{\partial z_{i\beta}} = 0.
\label{30}
\end{equation}

Therefore, there must exist 4-vectors $A_{i}^{\mu}$ such that:

\begin{equation}
F_{i}^{\mu\nu}(z_{i}) = \frac{\partial A_{i}^{\nu}}{\partial z_{i\mu}} - \frac{\partial A_{i}^{\mu}}{\partial z_{i\nu}}.
\label{31}
\end{equation}

Indeed, from (\ref{27}) and (\ref{31}) it is not difficult to verify that the 4-vectors $A_{i}^{\mu}$ can be written as:

\begin{equation}
A_{i}^{\mu}(z_{i}) = - \alpha\sum\limits_{j \ne i}^{} g_{j} \int ds_{j} \delta\left(\left(z_{i} - z_{j}\right)^{2}\right) \left(\dot{z}_{j} (z_{i} - z_{j})\right)
\left(z_{i}^{\mu} - z_{j}^{\mu}\right).
\label{32}
\end{equation}

Notice that from (\ref{32}) it follows that:

\begin{equation}
\frac{\partial A_{i}^{\nu}}{\partial z_{i}^{\nu}} = 0,
\label{a}
\end{equation}

\begin{equation}
\square^{2} A_{i}^{\mu}(z_{i}) = 8\pi \alpha \sum\limits_{j \ne i}^{} g_{j} \int ds_{j} \delta^{4} \left(z_{i} - z_{j}(s_{j})\right) \dot{z}_{j}^{\mu}(s_{j}).
\label{33}
\end{equation}

This immediately suggests that in a field theoretical formulation the field associated with a particle of charge $g$ following the world line $z(s)$ 
can be written as:

\begin{equation}
A^{\mu}(x) = - \alpha g \int ds \delta\left(\left(x - z(s)\right)^{2}\right) \left(\dot{z}(s) (x - z(s))\right)
\left(x^{\mu} - z^{\mu}(s)\right).
\label{34}
\end{equation}

Therefore, the field theory corresponding to the action-at-a-distance theory described by the action functional (\ref{25}) is governed by the equations:

\begin{equation}
m \ddot{z}^{\mu} = \frac{g}{c^{2}} F^{\mu\nu} \dot{z}_{\nu},
\label{35}
\end{equation}

\begin{equation}
\square^{2} A^{\mu} - \partial^{\mu} \left(\square \partial_{\nu} A^{\nu}\right) = 8\pi \alpha g \int ds \delta^{4} \left(x - z(s)\right) \dot{z}^{\mu}(s).
\label{36}
\end{equation}

Notice that $A^{\mu}$ given by equation (\ref{34}) obeys the Lorenz condition:

\begin{equation}
\partial_{\nu} A^{\nu} = 0.
\label{b}
\end{equation}

Therefore, $A^{\mu}$ given by (\ref{34}) is only one possible solution of (\ref{36}).

The theory described by the equations (\ref{35}) and (\ref{36})) was proposed in the 70's by Kiskis \cite{kiskis} as a model
for the strong interactions between quarks. 

It is not difficult to verify that the equations (\ref{35}) and (\ref{36}) can be derived from the following action functional:

\begin{equation}
S = -mc\int ds (\dot{z}^{2})^{\frac{1}{2}} - \frac{g}{c}\int ds A_{\mu} \dot{z}^{\mu} + \frac{1}{32\pi \alpha c}\int d^{4}x \left(\partial_{\sigma}F_{\mu\nu}\right)
\left(\partial^{\sigma}F^{\mu\nu}\right).
\label{37}
\end{equation}

This completes our construction of the field theory for the case (a) ($\alpha \ne 0, \beta=\gamma=0$) in (\ref{24}).

As can be seen from (\ref{36}, \ref{37}) this is an example of a field theory with a Lagrangian depending on higher derivatives, which
leads to complications\cite{barut}. Of the three cases considered, this is the one that appears in 
a more complex form when considered as a field theory. Nevertheless, it is interesting to note that this case is the simplest of the three cases considered 
when viewed as an action-at-a-distance theory. This can be seen by comparing (\ref{27}) with (\ref{7}) and (\ref{18}).
Notice that $F_{i}^{\mu\nu}$ given by formula (\ref{27}) does not depend on the 4-vector accelerations  $\ddot{z}_{j}$. 

It can be said that there is a sort of duality between the field formulation and the action at a distance formulation of a theory. 
In case (a) ($\alpha \ne 0, \beta=\gamma=0$) it appears that the action-at-a-distance formulation of the theory has a simpler form than the field formulation.

It is generally believed that strong interactions are best described by quantum chromodynamics, which is a field theory of the Yang-Mills type with $SU(3)$ as its
internal symmetry group. At the classical level the Yang-Mills fields $A_{a}^{\mu}$ ($a=1,...,d$), where $d$ is the dimension of the compact Lie group (the internal
symmetry group), obey the equations\cite{rubakov, kosyakov}:

\begin{equation}
\partial_{\nu} F_{a}^{\mu\nu} - g f_{abc} F_{b}^{\mu\nu} A_{c\nu} = - \frac{4\pi}{c} j_{a}^{\mu},
\label{38}
\end{equation}

\noindent where,

\begin{equation}
F_{a}^{\mu\nu} = \partial^{\mu} A_{a}^{\nu} - \partial^{\nu} A_{a}^{\mu} + g f_{abc} A_{b}^{\mu} A_{c}^{\nu}.
\label{39} 
\end{equation}

The structure constants $f_{abc}$ are real and completely antisymmetric.

It is well known that the Yang-Mills equations can be derived from the action functional:

\begin{equation}
S = - \frac{1}{c^{2}}\int ds A_{a\mu} j_{a}^{\mu} - \frac{1}{16\pi c}\int d^{4}x F_{a\mu\nu}F_{a}^{\mu\nu}.
\label{40}
\end{equation}

As a first step in an attempt to find an action-at-a-distance formulation of this theory, it may be convenient to assume the source to be a point particle
with world line $z^{\mu}(s)$ and internal degrees of freedom $q_{a}(s)$ ($a=1,...,d$) obeying Wong's equations\cite{wong,kosyakov}:

\begin{equation}
m \ddot{z}^{\mu} = \frac{1}{c^{2}} q_{a} F_{a}^{\mu\nu} \dot{z}_{\nu},
\label{41}
\end{equation}

\begin{equation}
\dot{q}_{a} = \frac{g}{c} f_{abc} q_{b} A_{c}^{\nu} \dot{z}_{\nu}.
\label{42}
\end{equation}

It is assumed here that the currents $j_{a}^{\mu}$ in (\ref{40}) take the form:

\begin{equation}
j_{a}^{\mu} = c \int ds \delta^{4} \left(x - z(s)\right) q_{a}(s) \dot{z}^{\mu}(s).
\label{43}
\end{equation}

The internal degrees of freedom $q_{a}$ ($a=1,...,d$) are assumed to be the components of an isovector under the adjoint representation of the symmetry group.

A conformally invariant version of the direct interparticle action for systems of massless particles with Yang-Mills interactions was studied in\cite{kosyakov1}.

In an action-at-a-distance description of Yang-Mills interactions the equations of motion of the particles may be postulated to be as follows:

\begin{equation}
m_{i} \ddot{z}_{i}^{\mu} = \frac{1}{c^{2}} q_{ia} F_{ia}^{\mu\nu} \dot{z}_{i\nu},
\label{44}
\end{equation}

\begin{equation}
\dot{q}_{ia} = \frac{g}{c} f_{abc} q_{ib} A_{ic}^{\nu} \dot{z}_{i\nu}.
\label{45}
\end{equation}

\noindent where,

\begin{equation}
F_{ia}^{\mu\nu} = \frac{\partial A_{ia}^{\nu}}{\partial z_{i\mu}} - \frac{\partial A_{ia}^{\mu}}{\partial z_{i\nu}} + g f_{abc} A_{ib}^{\mu} A_{ic}^{\nu},
\label{46} 
\end{equation}

\noindent with the condition that the functions $A_{ia}^{\mu}$ obey the equations:

\begin{equation}
\frac{\partial F_{ia}^{\mu\nu}}{\partial z_{i\nu}} - g f_{abc} F_{ib}^{\mu\nu} A_{ic\nu} = 
- 4\pi \sum\limits_{j \ne i}^{} \int ds_{j} \delta^{4} \left(z_{i} - z_{j}(s_{j})\right) q_{ja}(s_{j}) \dot{z}_{j}^{\mu}(s_{j}).
\label{47}
\end{equation}

A major difficulty in this approach is, of course, the nonlinearity of the equations (\ref{47}). Nevertheless, in this formulation the two-body problem
may admit exact circular orbits solutions, analogous to Schild's solutions\cite{schild} of the electromagnetic two-body problem. 

The main results presented in this paper are (i) a proof of a correspondence between the action-at-a-distance and field formulations 
of a relativistic theory of vector confinement given by the action functionals (\ref{25}) and (\ref{37}) and 
(ii) a proposal of a relativistic action-at-a-distance formulation of Yang-Mills interactions, which may simplify the computations when
searching for exact classical solutions of the two-body problem in this model.

\end{document}